\documentclass[a4paper, 11pt]{article} 
\usepackage{fullpage} 
\usepackage[utf8]{inputenc}
\usepackage[super,sort&compress]{natbib}
\setcitestyle{comma,numbers,super,open={},close={}} 
\makeatletter
\renewcommand\@biblabel[1]{#1.}
\makeatother
\usepackage{graphicx}
\usepackage{amssymb}
\usepackage{bbm,xcolor,color}
\usepackage{lscape}
\usepackage{amsmath, amsthm}
\usepackage[T1]{fontenc}
\usepackage[onehalfspacing]{setspace}
\usepackage{sectsty}
\usepackage{multicol}
\usepackage{url}
\usepackage{caption}

\makeatletter
\def\@xfootnote[#1]{%
  \protected@xdef\@thefnmark{#1}%
  \@footnotemark\@footnotetext}
\makeatother

\title{The BPgWSP test: a Bayesian Weibull Shape Parameter signal detection test for adverse drug reactions}

\begin{document}
	
\maketitle

\begin{center}
\large{Julia Dyck\footnote{Department of Business Administration and Economics, Bielefeld University,
Universitätsstraße 25, Bielefeld, 33615, Germany.}\footnote[*]{Corresponding author: j.dyck@uni-bielefeld.de} and Odile Sauzet\footnotemark[1]\footnote{School of Public Health, Bielefeld University,
Universitätsstraße 25, Bielefeld, 33615, Germany.}
}  
\end{center}

\medskip

\begin{abstract}
We develop the Bayesian Power generalized Weibull shape parameter (BPgWSP) test as statistical method for signal detection of possible drug-adverse event associations using electronic health records for pharmacovigilance. The Bayesian approach allows the incorporation of prior knowledge about the likely time of occurrence along time-to-event data.
The test is based on the shape parameters of the Power generalized Weibull (PgW) distribution. When both shape parameters are equal to one, the PgW distribution reduces to an exponential distribution, yielding a constant hazard function. This is interpreted as no temporal association between drug and adverse event (AE). The BPgWSP test involves comparing a region of practical equivalence (ROPE) around one reflecting the null hypothesis with estimated credibility intervals (CI) reflecting the posterior means of the shape parameters. The decision to raise a signal is based on the CI+ROPE tests and the selected combination rule for these outcomes.

The test development requires a simulation study for tuning of the ROPE and CIs to optimize specificity and sensitivity of the test. Samples are generated under various conditions, including differences in sample size, prevalence of adverse drug reactions (ADRs), and the proportion of AEs. We explore prior assumptions reflecting the belief in the presence or absence of ADRs at different points in the observation period. Various types of ROPE, CIs, and combination rules are assessed, and optimal tuning parameters are identified based on the area under the curve.
The tuned BPgWSP test is illustrated in a case study in which the time-dependent correlation between the intake of bisphosphonates and four AEs is investigated.

\end{abstract}


\section{Introduction}

In the field of pharmacovigilance, signal detection methods are used  to identify new potential adverse drug reactions\citep{EMAwebsiteADR} (ADRs) among adverse events \citep{EMAwebsiteAE} (AEs) occurring in the population by testing for an association between the occurrence of an AE and a prescribed drug.\citep{meyboom1997pharmacovigilance}
Signal detection is important to keep a drug‘s harm profile updated and can result in adjustments of the prescription labelling or even a recall of the product from the market. \citep{hauben2005, noren2010temporal, sauzet2022}
\\

With the collection of longitudinal electronic health records\citep{longEHR, epidatabases} (EHRs) (regularly collected data about diagnoses, treatments, and medication prescribed by physicians, as well as AEs), statistical methods to exploit them have emerged.
These include disproportionality approaches adjusted for longitudinal data comparing rates of occurrence between a single drug and the rest of the dataset \citep{zorych2013disproportionality,schuemie2011lgps}, temporal pattern analysis \citep{noren2010temporal}, exposure models \citep{exposure_models2024preprint} and more (see for example the review by Coste et al. \cite{coste2023methodreview}).

Sauzet and Cornelius\cite{cornelius2012, sauzet2022} developed a family of signal detection tests based on an assumption about time dependence on the drug prescription.
More specific, their tests aim for detection of a non-constant hazard function (depicting the risk over time) by modelling right-censored time-to-event data starting at drug prescription to flag a possible drug-event association.  
Making assumptions about the generating process of an outcome (more specifically the hazard function) in the development of a test of association can improve the performance of this test in terms of power, specificity, and sensitivity by enabling the use of more statistical information included in the data.
However, this is not the only type of information that can be incorporated into a test. 

Prior knowledge regarding the distribution parameters of the generating process (e.g. mean time of occurrence of an event) might be incorporated into the development of a test using a Bayesian approach. 
Bayesian statistical methods like the Empirical Bayes Gamma Mixture, also called Gamma-Poisson shrinker, or the Bayesian confidence propagation neural network are extensions of disproportionality measures used to obtain signals from spontaneous reporting databases.\citep{bansal2025, PrietoMerino2011BayesianPharmacovigilance} These, in turn, do not consider the time of the event occurrence.  
\\

This article aims to combine the assumption of a temporal process and the incorporation of prior knowledge about distribution parameters of this process.
To do so, we will generalize the power generalized Weibull shape parameter (pWSP) test proposed by Sauzet and Cornelius \cite{sauzet2022} into a Bayesian framework to develop and tune an HDI+ROPE\cite{kruschke2018} type test (a Bayesian equivalent of a hypothesis test) of signal detection which we call the Bayesian Power generalized Weibull shape parameter (BPgWSP) test. 
\\

\section{Methodological concept for the Bayesian signal detection test}\label{sec:method}

\begin{figure}
    \centering
    \includegraphics[width=0.5\linewidth]{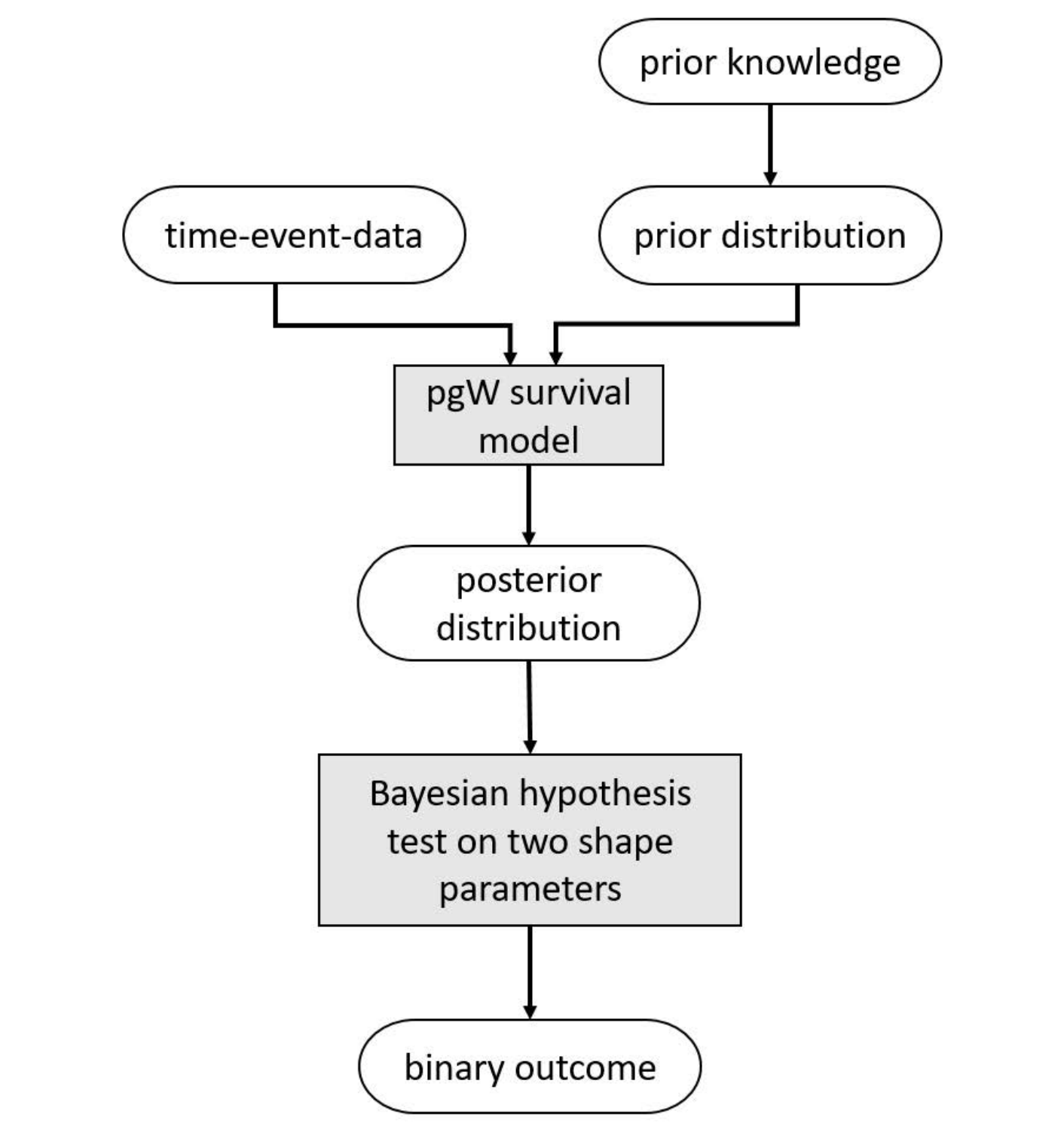}
    \caption{Methodological concept for the signal detection test. }
    \label{fig:concept_signal_detection_test}
\end{figure}

The principle of non-constant hazard signal detection tests is built on the assumption that the risk of an AE is time-dependent relative to the prescription of the drug under investigation.

The method is conceptualized for right-censored time-to-event data. Each observation $i$ consists of the time $t_i \in (0, c)$ with $c$ being the censoring time and the binary status $d_i \in \{0,1\}$ indicating whether the AE occurred at $t_i$.
The parametric distribution is defined by a parameter vector $\Theta$. 
We define the null-hypothesis of constant hazard based on a subset of parameters $\mathcal{N} \subset \Theta$ and develop the Bayesian signal detection test (see Figure \ref{fig:concept_signal_detection_test}) based on an updated distribution for $\mathcal{N}$ using the HDI+ROPE approach (see sec. \ref{sec:method_for_dev_of_Badr_test}). For development, we will perform the three steps below:

\begin{enumerate}
    \item present the power generalized Weibull (PgW) distribution which we will use in this paper and formulate the null-hypothesis,
    \item describe the BPgWSP test using the HDI+ROPE approach based on a Bayesian estimation,
    \item set the prior specification and HDI+ROPE tuning parameters for the BPgWSP test by performing a simulation study (see sec. \ref{sec:simulation_study}).
\end{enumerate}

\subsection{Power generalized Weibull distribution}
The survival function of the PgW distribution\cite{bagdonavicius2001, nikulin2016} is
\begin{align}
S(t) = \text{exp}\left\{ 1 - \left[ 1 + \left(\frac{t}{\theta}\right)^{\nu}\right]^{\frac{1}{\gamma}} \right\}.
\end{align}
The hazard function\cite{bagdonavicius2001, nikulin2016} is
\begin{align}
h(t) = \frac{\nu}{\gamma \theta^\nu}\cdot t^{\nu-1} \cdot \left[1 + \left(\frac{t}{\theta}\right)^{\nu}\right]^{\frac{1}{\gamma} - 1} .
\end{align}

The PgW distribution has three parameters $\Theta_{PgW} = \{\theta, \nu, \gamma\}$ - two of which are shape parameters such that $\mathcal{N}_{PgW} = \{\nu, \gamma\}$.
\begin{figure}[t]
    \centering
    \includegraphics[width = \textwidth]{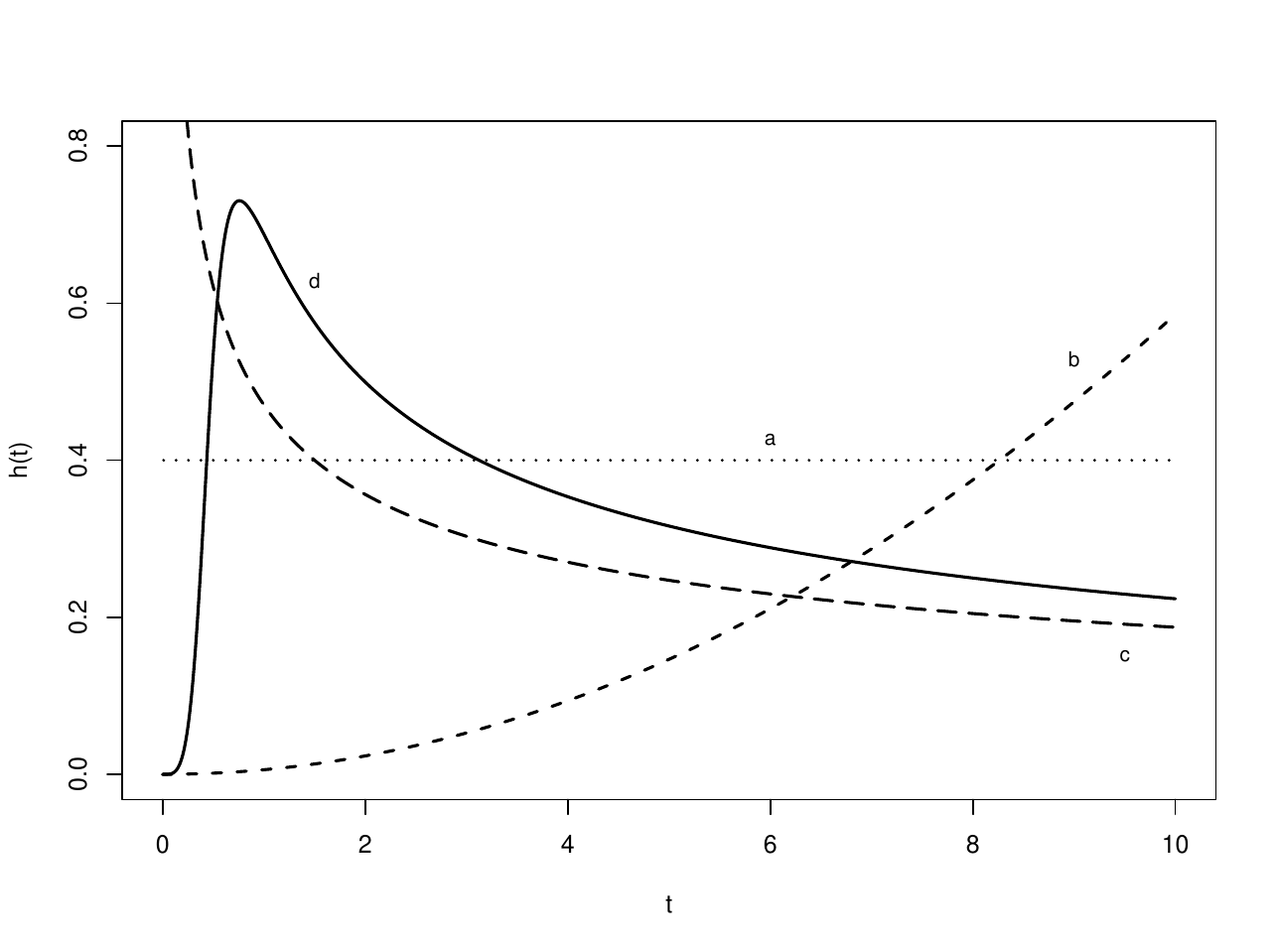}
    \caption{Hazard functions obtained from power generalized Weibull distributions with a) constant ($\theta = 2, \nu = 1, \gamma = 1$) , b) increasing ($\theta = 8, \nu = 3, \gamma = 1$), c) decreasing ($\theta = 1.5, \nu = 0.6, \gamma = 1$), d) unimodal ($\theta = 0.5, \nu = 5, \gamma = 10$), e) bathtub ($\theta = 30, \nu = 0.9, \gamma = 0.2$) form.}
    \label{fig:pgwforms}
\end{figure}
The two shape parameters allow for a range of hazard function forms including constant, monotoneously in- or decreasing, and unimodal or bathtub shaped (see Figure \ref{fig:pgwforms}). 

A constant hazard is given if and only if both shape parameters are equal to one, so that the null-hypothesis of constant hazard can be formulated as

\begin{align}\label{eq:null_hypothesis}
    H_0: \nu = 1 \text{ and } \gamma = 1 .
\end{align}

\subsection{Development of Bayesian signal detection test using the HDI+ROPE approach}\label{sec:method_for_dev_of_Badr_test}


The prior distribution of $\Theta$ should reflect whether the AE of interest is suspected to be an ADR (by choosing a parameter combination yielding a non-constant hazard) and if so, when it is expected to occur a priori by medical experts (by specifying parameters $\Theta$ that lead to the a priori intended $\mathbf{E}[t]$). 
The scale parameter reflects the AE background rate. The shape parameters $\mathcal{N} \subset \Theta$ determine the constancy or non-constancy and general trend of the hazard over time.

For practicality, univariate prior distributions for each parameter out of $\Theta$ are specified. Concerning implementation, not $\mathbf{E}[t]$ but the prior means of all parameters are set, followed by checking whether the hazard form and $\mathbf{E}[t]$ resulting from the chosen prior means reflect the prior belief to a sufficient extent in a trial-and-error manner.

Based on the specified prior means (and a chosen standard deviation) as well as a distributional choice (e.g. the product of univariate lognormal distributions for each parameter in $\Theta$), the prior distribution $p_0(\Theta)$ will be updated yielding a posterior distribution.
For time-to-event data the posterior distribution $p_1(\Theta|t)$ is proportional to
\begin{align}
p_1(\Theta|t) \propto p_0(\Theta)\cdot \mathcal{L}(t| \Theta)
\end{align}
with
\begin{align}
    \mathcal{L}(t| \Theta) = \prod_{i=1}^N S(t_i)^{1-d_i}\cdot f(t_i)^{d_i}
\end{align}
 being the likelihood function in the time-to-event framework such that all subjects $i$ not having an AE during the observation time (denoted by $d_i = 0$) are considered in terms of the survival function $S(t_i)$ and all $i$ experiencing an AE (denoted by $d_i = 1)$ are taken into account in terms of the density function $f(t_i) = S(t_i)\cdot h(t_i)$.
\\

Based on the posterior distribution of $\mathcal{N}$, the HDI+ROPE test \citep{kruschke2015, kruschke2018} is applied to each shape parameter and combined, leading to two possible outcomes: either we lean towards the null-hypothesis of constant hazard or the alternative hypothesis of a non-constant hazard. The test setup depends on 
\begin{itemize}
    \item  the region of practical equivalence (ROPE) to the null value (in our case, one), typically defined as an interval around the null value 
    and
    \item a credibility interval (CI) based on posterior distribution parameters in  $\mathcal{N}$, here either a highest density interval (HDI) as proposed by Kruschke\cite{kruschke2015} or an equal-tailed interval (ETI). See sec. \ref{sec:tuningpars} for the formal definition of the ETI and HDI.
\end{itemize}

\begin{figure}
    \centering
    \includegraphics[width=\linewidth]{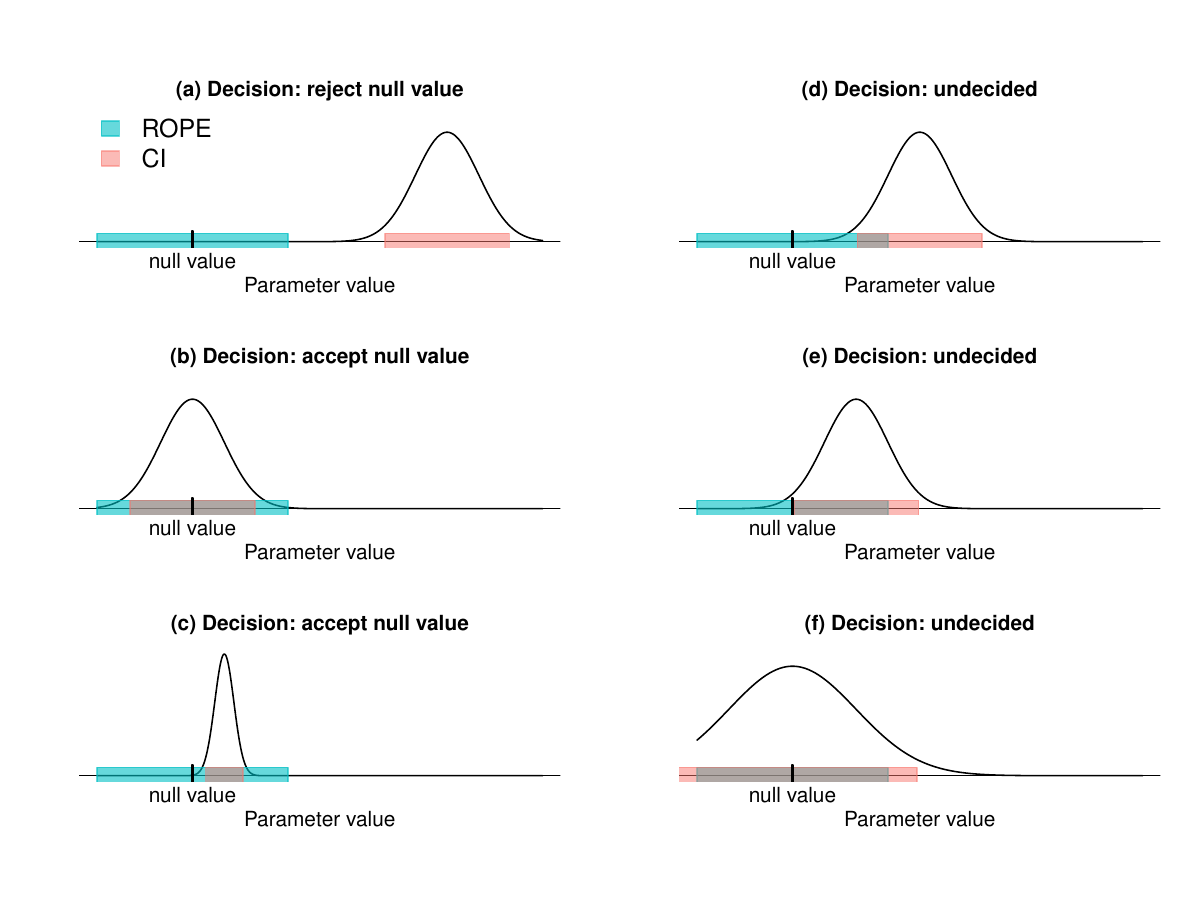}
    \caption{HDI+ROPE decision rules based on graphic from Kruschke, Fig. 1.\cite{kruschke2018}}
    \label{fig:HDI+ROPE_outcomes}
\end{figure}

In our setting, the HDI+ROPE test is applied to each parameter $\nu \in \mathcal{N}$ and provides one of the following interim results:
\begin{itemize}
    \item If the CI lies completely within the ROPE, then the single parameter null-hypothesis$$H_0(\nu): \nu = 1$$ is accepted (Figure \ref{fig:HDI+ROPE_outcomes}b-c),
    \item if ROPE and CI do not intersect at all, then the single parameter null-hypothesis is rejected (Figure \ref{fig:HDI+ROPE_outcomes}a),
    \item if ROPE and CI partly overlap, no decision is made (Figure \ref{fig:HDI+ROPE_outcomes}d-f).
\end{itemize}
Given the single parameter test outputs, the combined test result is concluded according to a combination rule, for instance, the condition implied by Equation \ref{eq:null_hypothesis}:
\begin{itemize}
    \item If the HDI+ROPE accepts the null-hypothesis for all parameters in $\mathcal{N}$, then accept the null-hypothesis of constant hazard.
    \item Else, reject the null-hypothesis of constant hazard and raise a signal.
\end{itemize}

To complete the test development, we need to make decisions about the prior distributional choice, the position and width of the ROPE and the CI of the BPgWSP test and the combination rule. 

\section{Simulation study for test development and tuning}\label{sec:simulation_study}

A method to finalize the BPgWSP test, theoretically described in sec. \ref{sec:method}, is to perform a simulation study.
By applying a range of prior distributions, choices for position and width of the ROPE, position and width of the CI, and combination rules, we make decisions about the optimal tuning parameters for the BPgWSP test such that a performance parameter based on sensitivity and specificity of the test is maximal across a range of sample scenarios.
The frequentist pWSP test approach developed by Sauzet and Cornelius \cite{sauzet2022} with credibility levels to be tuned is considered in the simulations for comparison reasons.

\subsection{Data generating process}\label{sec:datagenpars}

The sample scenarios are determined by the
\begin{itemize}
    \item censoring time (365 days) setting the end of the observation period (OP),
    \item sample size $N$ (500, 3000, 5000),
    \item backround rate (0.1) which specifies the prevalence in the population,
    \item ADR rate which is the proportion of the background rate (0, 0.5, 1),
    \item expected event-time of the ADR if the ADR rate is above zero (1st, 2nd, 3rd quarter of OP),
    \item relative standard deviation of the ADR event-time (0.05).

\end{itemize}

The background observations are sampled from a binomial distribution with a probability equal to the specified background rate
$$ n^{br} \sim Bin(n = N, p = br). $$
Corresponding event-times are sampled from a uniform distribution over the OP ranging from day one to $365$.
\begin{align}
     t^{br} & \sim Unif(1,365)
\end{align}
The number of events induced by the drug is sampled from a binomial distribution with the probability being the background rate times ADR rate. 
\begin{align}
    n^{ADR} \sim Bin(n = N, p = br \cdot adr)
\end{align}
Event-times for the ADR cases are generated from a normal distribution with an expected event-time $\mathrm{E}[t]$ at the first, second, and third quarter of the OP and a standard deviation of $0.05$ times the censoring time at 365 days.
\begin{align}
t^{ADR} & \sim \mathcal{N}(\mathrm{E}[t], (0.05 \cdot 365)^2)\
\end{align}

The sample is filled up to size $N$ with censored observations. They have the AE status zero and time point $365$ (end of the OP).

\subsection{Prior specifications and model estimation}\label{sec:priorspec}

 \begin{figure}[t]
    \centering
    \includegraphics[width = \textwidth]{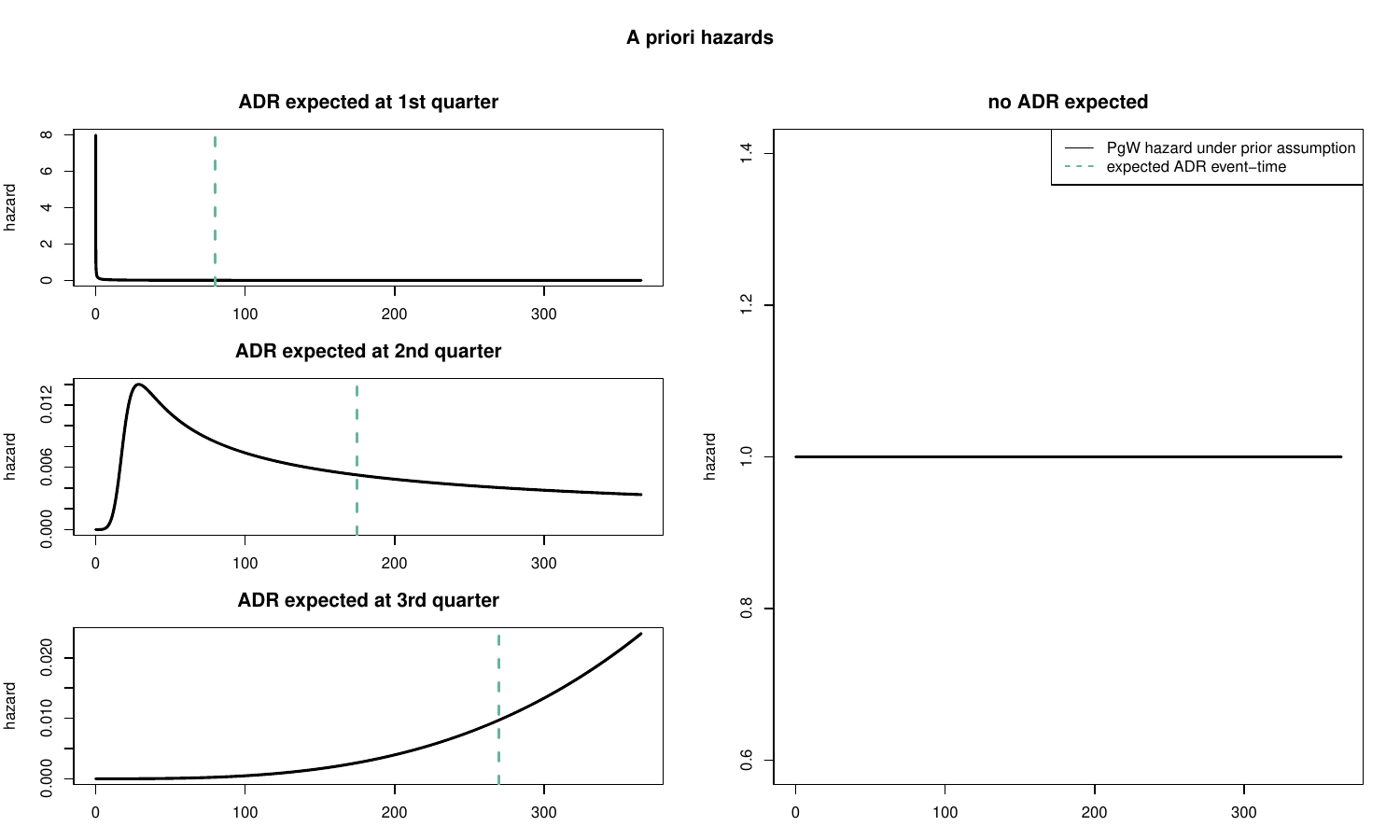}
    \caption{Hazard functions obtained under the prior belief that the expected event-time is at the first, second, or third quarter of the OP (left) or that the AE of interest is no ADR (right).}
    \label{fig:priorforms}
\end{figure}


Four different prior beliefs are implemented. An ADR is expected to occur with higher risk either around the first, second, or third quarter of the OP or not expected at all. These assumptions are represented by hazard functions of the distribution shown in Figure \ref{fig:priorforms} based on chosen prior mean values for the scale $\theta$, and shapes $\nu$ and $\gamma$ listed in Table \ref{tab:sim_priorbelief}. The standard deviation of all parameters is set to 10. See the supplementary material for details on the implementation.
\\

\begin{table}[h]
    \centering
    \begin{tabular}{|l|r|c|}
        \hline
        prior belief about & a priori & prior means  \\
        expected event-time & $E[t]$ & $(\bar{\theta}_0, \bar{\nu}_0, \bar{\gamma}_0)$  \\
        \hline
        none & - & $(1, 1, 1)$    \\
        1st quarter of OP & 91 days & $(1, 0.207, 1)$  \\
        2nd quarter of OP  & 183 days & $(20, 5.5, 14)$  \\
        3rd quarter of OP  & 274 days & $(300, 4, 1)$ \\
        \hline
    \end{tabular}
    \caption{Mean of prior distribution under implemented prior beliefs about expected event-time. Prior standard deviation is set to 10 for all parameters in all prior belief cases. }
    \label{tab:sim_priorbelief}
\end{table}

To obtain the posterior estimate the No-U-Turn sampler is used \cite{NUTS2014}. We draw a posterior sample of size $40\,000$ obtained from four chains of length $10\,000$ after a burn-in phase of $1\,000$ draws per estimation. In practice, model diagnostics should be used to check convergence and sampling efficiency before using it for testing.\cite{stanmanual} This is not done within the simulations due to the automated procedure, but showcased in the supplementary material.
\\

Each combination of sample scenario and prior specification leads to a total of 168 simulation settings -- 144 with a positive ADR rate and 24 controls with only background events to match.
Each simulation scenario's data generation and posterior estimation is repeated 100 times. 

The pWSP test does not require a prior specification, such that the number of simulation scenarios is reduced to 21 -- 18 scenarios with a positive ADR rate and three controls. Data generation and maximum likelihood estimation are performed 800 times each.

\subsection{Specification of the HDI+ROPE test }\label{sec:tuningpars}

The Bayesian HDI+ROPE test approach is specified with respect to the
\begin{itemize}
    \item ROPE position and width,
    \item posterior CI position and width,
    \item combination of the single test outputs of both shape parameters.
\end{itemize}

We specify the ROPE for each shape parameter $\nu$ and $\gamma$ as equal-tailed interval (ETI) 
\begin{align}\label{eq:ETI}
    [q_{(1-\alpha)/2}, q_{(1+\alpha)/2}]
\end{align}
based on the quantiles  $q$ of the shape parameters' prior distributions under the null-hypothesis at credibility level $\alpha$.

The posterior CI at credibility level $\alpha$ for each shape parameter is set to be an ETI (see Equation \ref{eq:ETI}) calculated from the empirical quantiles of the shape parameters' posterior samples or an HDI  estimated based on the posterior sample.
An HDI at level $\alpha$ is an interval
\begin{equation}
    HDI(\nu) = \{\nu \; |\; p_1(\nu) \geq w\}\\ \text{ with } w\in [0,1] \text{ such that} \int_{\nu \; | \; p_1(\nu) \geq w} p_1(\nu|t)\;  d\nu = 1 - \alpha ,
\end{equation}
that is with boundaries such that all values inside the HDI have a higher density than values outside the HDI and simultaneously encompasses $(1-\alpha)\cdot100\%$ of the probability mass.\citep{kruschke2015}

To control the width of the ROPE as well as the posterior CIs, we apply credibility levels $\alpha$ ranging from $0.5,0.55,...0.95$.
\\
\begin{table}[h]
    \centering
\begin{tabular}{| c | c | c | c | c |}
\hline
 \multicolumn{2}{|c|}{} & \multicolumn{3}{c|}{}\\ 
				 \multicolumn{2}{|c|}{interim result}  & \multicolumn{3}{c|}{combination rule} \\ 
     \hline
				&  & option 1  & option 2 & option 3\\
				for $\nu$ & for $\gamma$ & (acc. to Eq. \ref{eq:null_hypothesis})  & (reserved) &(very reserved) \\
				\hline  
				rejected & rejected & signal & signal & signal \\   
				accepted & rejected & signal & - & - \\   
				rejected & accepted  & signal & - & - \\   
				accepted & accepted & - & - & - \\   
				undecided & rejected & signal & signal & - \\   
				undecided & accepted & - &- & - \\   
				rejected & undecided & signal & signal & - \\   
				accepted & undecided & - &  - & - \\   
				undecided & undecided & signal & - & - \\
    \hline
			\end{tabular}
    \caption{Three options for the combination of the single test outcomes for shape parameter $\nu$ and powershape parameter $\gamma$. Option 1 raises a signal if at least one parameter significantly differs from one. Option 2 raises a signal in fewer cases, namely when one parameter test rejects the null hypothesis $H_0$ and the other rejects $H_0$ or is undecided. Option 3 raises a signal only if both single parameter tests reject $H_0$.}
    \label{tab:combination_rules}
\end{table}

Three different combination rules of the single shape parameters' test outcomes are defined in Table \ref{tab:combination_rules}. Option 1 implements the hypothesis formulated in Equation \ref{eq:null_hypothesis}.
A signal is raised if at least one shape parameter significantly differs from one indicating a non-constant hazard function. Option 2 raises a signal in fewer cases, namely if the null-hypothesis is rejected for at least one shape parameter and rejected or undecided for the other. Option 3 raises a signal if and only if both single shape parameter tests reject the null-hypothesis.
\\

The pWSP test is constructed according to Sauzet and Cornelius.\cite{sauzet2022} Based on the maximum likelihood estimate (MLE) the confidence interval at confidence level $\alpha$

\begin{align}
    [  \hat{\nu}_{MLE} \pm z_{(1+\alpha)/2} \cdot sd(\hat{\nu}_{MLE})]
\end{align}

  is calculated and used for significance testing for each shape parameter. We apply confidence levels $0.5,0.55,...0.9$ and $0.91,...0.99$ as well as $0.991,...0.999$ to take into account the relevant levels according to Sauzet et al. \cite{Sauzet2024}
 When shape parameters cannot be estimated (e.g. due to convergence issues), this is considered as no rejection of the single parameter null-hypothesis. 

A signal is raised if both parameter estimates differ significantly from the null value.
\\

The simulation study is implemented in the statistical software environment R \citep{Rsoftware} 
including the packages rstan \citep{rstanpackage} 
for the Bayesian model setup and posterior sampling,  HDInterval \citep{HDIntervalpackage} 
for calculation of posterior CIs, and BWSPsignal \cite{BWSPsignal} containing functions for the BPgWSP test and simulation framework.

\subsection{Evaluation criteria for the simulation results}

The evaluation of the simulation study is performed in three steps.
First, prior distributions that show convergence issues, extremely high execution times, or small posterior effective sample sizes (ESS) are excluded. The model is considered to have not converged whenever the \verb|rstan::sampling| function returns an error. We consider the execution time extreme if the maximum is 10 hours or more per model. Finally, following Kruschke's recommendation for HDI+ROPE testing, we consider prior distributions exceeding a posterior ESS of $ 10,000$ or more on average as sufficient for testing. \cite{kruschke2015, stanmanual}

Second, the tuning parameter combinations are classified based on the test sensitivity and specificity under correct prior beliefs.
The performance of the implemented test setups averaged over varying simulation scenarios  is measured in terms of the area under the curve (AUC) of the receiver operating characteristic (ROC) graph with one threshold \citep{fawcett2004, lloyd1998} based on equal numbers of ADR and no-ADR scenarios. In addition to the AUC value, the ROC curve allows further investigation of the top tests by displaying the AUC depending on the false and true positive (FP, TP) rates.
The AUC is calculated using the R package ROCR.  \citep{ROCR}

Third, to explore the effect of sample size, ADR rate, and expected event-time on the test performance, we present grouped AUCs of the best test.
In case of a BPgWSP test, the robustness to miss-specification is explored by examining the AUC when the prior belief differs from the truth and the expected event-time is therefore incorrectly specified.

\subsection{Results of the simulation study}

We compare the prior distributions with respect to convergence issues, execution times, and ESSs. There were no convergence issues.

 \begin{figure}[p]
    \centering
    \includegraphics[width = 0.8\textwidth]{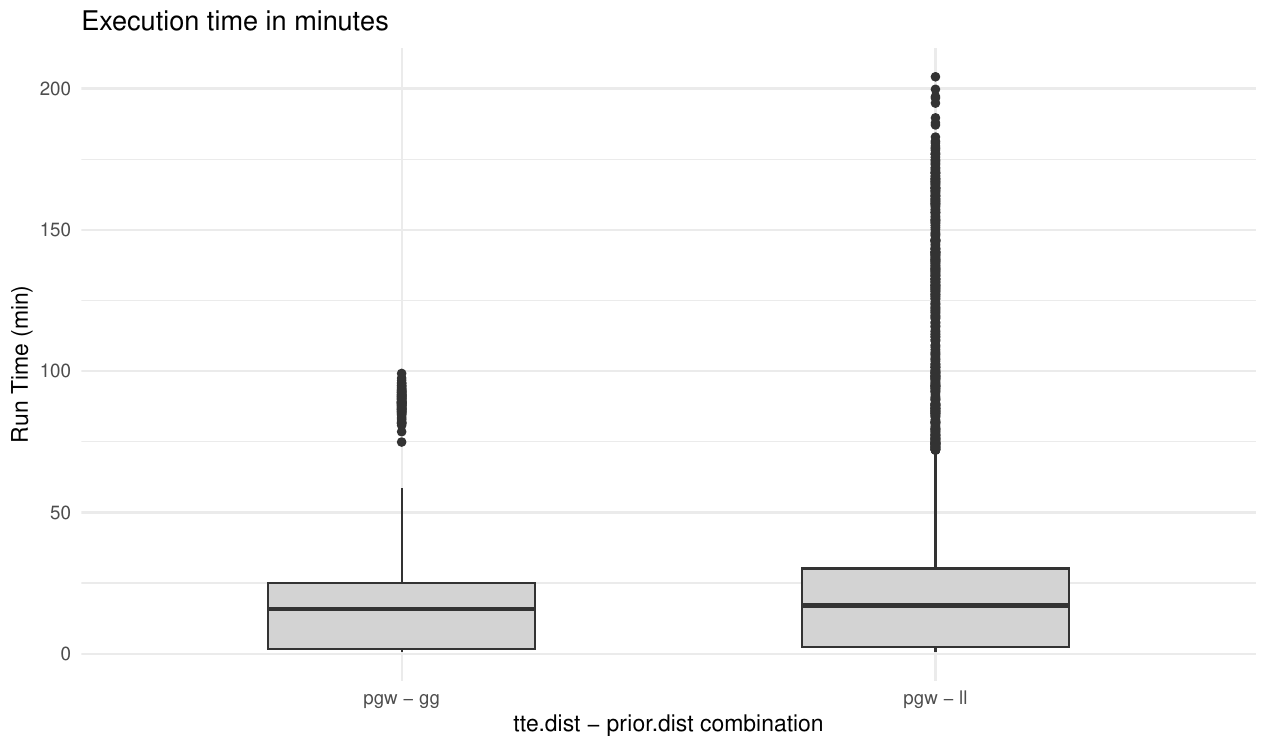}
    \caption{Boxplot comparing execution times in minutes for two prior distributions, gamma (pgw-gg) and lognormal (pgw-ll), with lognormal yielding higher maximum run times of up to 3 h, but both prior distributions producing similar median times of 16 Minutes with gamma and 17 Minutes with lognormal prior.}
    \label{fig:runningtimes}

    \centering
    \includegraphics[width = 0.8\textwidth]{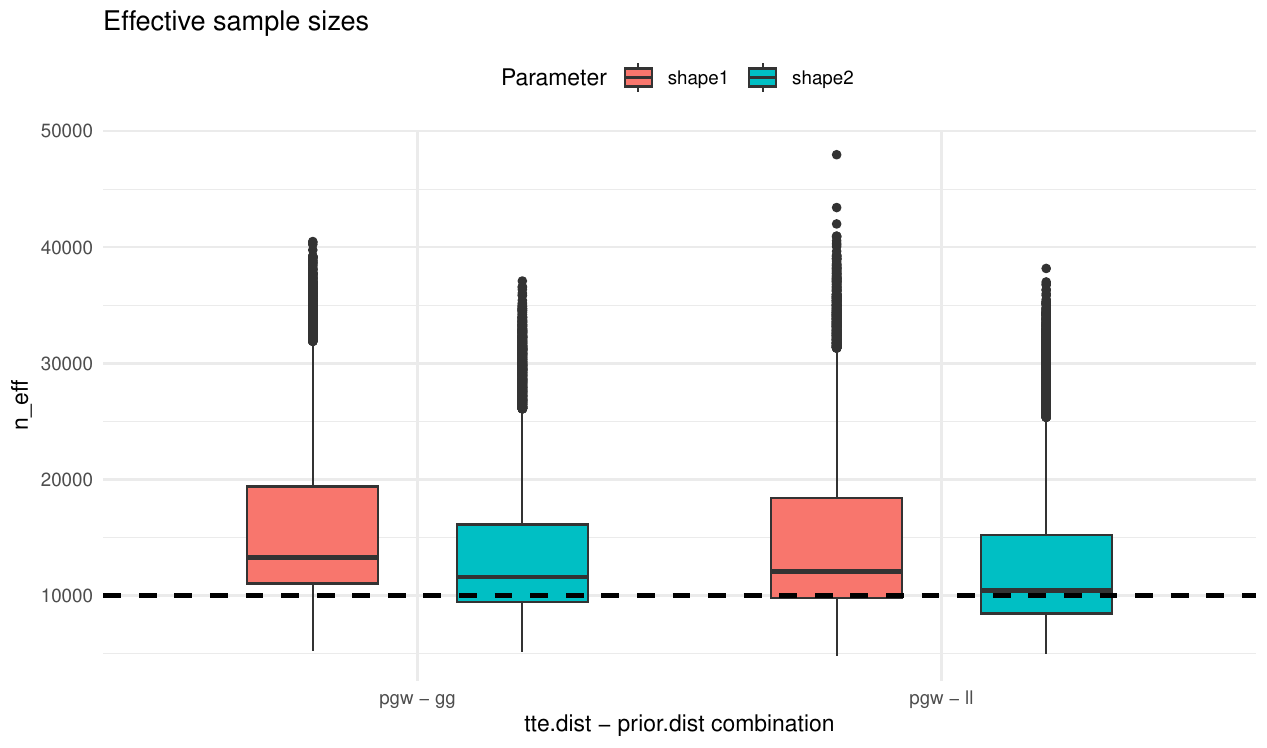}
    \caption{Boxplots of estimated effective sample sizes (n\_eff) split with respect to prior distributional choices gamma (pgw-gg) and lognormal (pgw-ll) for parameter $\nu$ (shape1 in red) and $\gamma$ (shape2 in blue).
    The recommended threshold for the effective sample size is marked with a horizontal dashed line at $10\,000$.
    }
    \label{fig:effectivesamplesize}
\end{figure}
A boxplot of the execution times for posterior estimation stratified with respect to the prior distribution is shown in Figure \ref{fig:runningtimes}. The models produce median execution times of 16 Minutes with gamma priors, and 17 Minutes with lognormal priors. 
The maximal running times  stayed under 1 h 40 min given a gamma prior and under 3h given a lognormal prior. 
Gamma priors lead to ESSs of $>10\,000$ in $85\%$ of the posterior estimations for $\nu$, and $68\%$ for $\gamma$.
For lognormal priors these number are $73\%$ for $\nu$ and $55\%$ for $\gamma$ (see Figure \ref{fig:effectivesamplesize}). 

Based on execution time and ESS results, both prior distribution choices are acceptable for the modeling part of the BPgWSP. 
\\

 \begin{figure}[t]
    \centering
    \includegraphics[width = \textwidth]{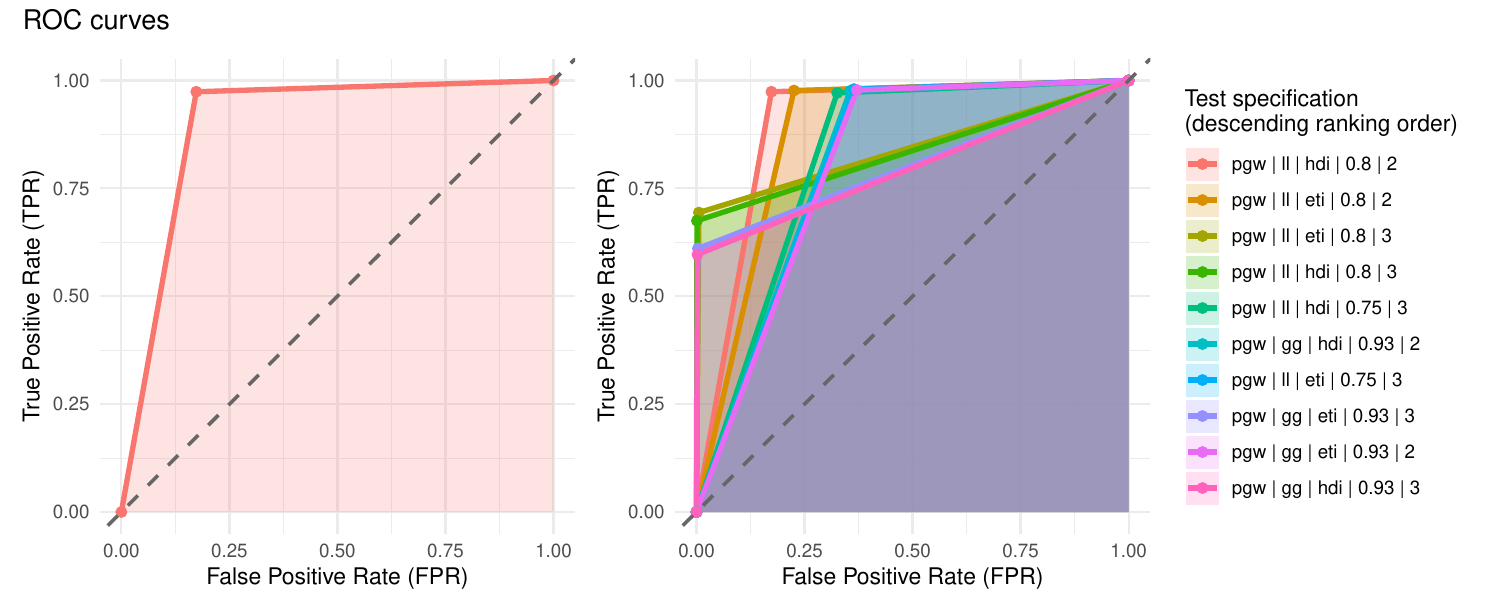}
    \caption{Roc curves representing the top one (left) and top ten (right) BPgWSP tests' performances.
    }
    \label{fig:roccurves}
\end{figure}

\begin{table}[h]
\centering
\begin{tabular}{|r|llllrrr|}
\hline
rank & prior  & ROPE & posterior CI & combination  & AUC & FP rate & TP rate \\
 &  distribution & &  &  rule &  &  &  \\
\hline
1 & lognormal & 80\% ETI & 80\% HDI & option 2 & 0.900 & 0.173 & 0.974 \\
2 & lognormal & 80\% ETI & 80\% ETI & option 2 & 0.876 & 0.226 & 0.977 \\
3 & lognormal & 80\% ETI & 80\% ETI & option 3 & 0.844 & 0.006 & 0.694 \\
4 & lognormal & 80\% ETI & 80\% HDI & option 3 & 0.837 & 0.001 & 0.675 \\
5 & lognormal & 75\% ETI & 75\% HDI & option 3 & 0.823 & 0.326 & 0.971 \\
6 & gamma & 93\% ETI & 93\% HDI & option 2 & 0.810 & 0.356 & 0.976 \\
7 & lognormal & 75\% ETI & 75\% ETI & option 3 & 0.808 & 0.364 & 0.980 \\
8 & gamma & 93\% ETI & 93\% ETI & option 3 & 0.804 & 0.003 & 0.611 \\
9 & gamma & 93\% ETI & 93\% ETI & option 2 & 0.803 & 0.371 & 0.978 \\
10 & gamma & 93\% ETI & 93\% HDI & option 3 & 0.797 & 0.002 & 0.597 \\
\hline
\end{tabular}
\caption{Top ten BPgWSP and pWSP test specifications ranked based on AUC. AUC, FP rate and TP rate are averaged over all sample scenarios given a correct prior belief.}
\label{tab:auc_ranking}
\end{table}
Table \ref{tab:auc_ranking} shows the top ten test specifications in terms of AUC along with the corresponding FP and TP rates. Figure \ref{fig:roccurves} displays the ROC curves of the top one (left) and top ten (right) test specifications.
The highest AUC value of $0.9$ is achieved with
an $80\%$ lognormal ETI as ROPE, an $80\%$ posterior HDI as posterior CI and option 2 for the combination rule. The corresponding ROC curve exhibits a TP rate of $0.97$ and a FP rate of $0.17$, indicating a trade-off that prioritizes the detection of TPs. 
The best pWSP test (frequentist) with confidence level of $55\%$ is ranked in 11th place with an AUC of $0.75$ under the given simulation conditions.

The influence of the sample size $N$, the proportion of ADR events and the expected event-time of the ADR on the AUC under a correctly specified prior belief is summarized in Table \ref{tab:effect_of_framework_cond}. The BPgWSP test works better with increasing $N$ reaching an AUC of $0.978$ given an $N$ of 5000.
An increase in the proportion of ADR events from $5\%$ to $10\%$ of $N$ increases the AUC very slightly from $0.888$ to $0.913$.
The test performance is higher for ADRs occurring around the first or third quarter of the OP with an AUC of $0.969$ or $0.946$, respectively, but still provides an AUC of $0.786$ for ADRs occurring in the middle of the OP.
\begin{table}[h!]
\centering
\begingroup\normalsize
\begin{tabular}{|lr|}
  \hline
grouping factors& AUC \\ 
\hline
   \multicolumn{2}{|l|}{\textbf{sample size $N$}} \\
    $500$ & 0.789 \\ 
    $3000$ & 0.933 \\ 
    $5000$ & 0.978 \\ 
   \hline
   \multicolumn{2}{|l|}{\textbf{proportion of ADR events}} \\
  $5\%$ of $N$ & 0.888 \\ 
    $10\%$ of $N$ & 0.913 \\ 
   \hline
   \multicolumn{2}{|l|}{\textbf{expected time of ADR events}} \\
   1st quarter of observation period & 0.969 \\ 
   2nd quarter of observation period & 0.786 \\ 
   3rd quarter of observation period & 0.946 \\ 
   \hline
\end{tabular}
\endgroup
\caption{Average AUC performance of the BPgWSP test with lognormal prior distribution, $80\%$ lognormal confidence interval as ROPE,  $80\%$ HDI as posterior CI, and combination rule option 2 grouped by sample size, proportion of ADR events, and expected time of ADR events given a correct prior belief.} 
\label{tab:effect_of_framework_cond}
\end{table}

If the specification of the expected event-time is off by one quarter of the observation time, the AUC reduces to $0.783$ on average. If the a priori expected event-time is off by two quarters, the AUC performance drops to $0.489$. An AUC performance of $0.787$ is provided in case no ADR is suspected a priori (see Table \ref{tab:robustness}).
\begin{table}[h!]
    \centering
    \begin{tabular}{|l|c|}
        \hline
        type of prior belief & AUC  \\
         \hline
        correct specification & 0.900 \\
        one quarter off & 0.783 \\
        two quarters off & 0.489 \\
        no ADR assumed & 0.787 \\
        \hline
    \end{tabular}
    \caption{AUC performance given a correct or incorrect prior belief of different types about the expected event-time.}
    \label{tab:robustness}
\end{table}
\\

For application, we recommend setting the tuning parameters as follows: an $80\%$ lognormal ETI for the ROPE, an $80\%$ posterior HDI for the posterior CI and option 2 for the combination rule.

\section{Case study}\label{sec:casestudy}

The purpose of this case study is to illustrate the BPgWSP test with a real dataset. Conclusions from this section should not be
used to inform clinical practice.
The dataset was originally created from the Health Improvement Network (THIN) database \citep{THINwebsite} to assess the association between the prescription of bisphosphonates and the occurrence of carpal tunnel syndrome in menopausal women.\citep{carvajal2016carpal} The THIN database contains primary care data from patients in the UK.

The BPgWSP test is applied to investigate the association between bisphosphonates and four AEs occurring in the population. 
In addition to the exposed cohort, the dataset contains patients not exposed to bisphosphonates (control group). A control group is not required to apply the BPgWSP test, but allows to explore potential type I errors.
The data is described in more detail in Sauzet et al.  \cite{sauzet2013illustration}

We illustrate the BPgWSP test using four different prior beliefs for the lognormal parameters reflecting an a priori expected ADR event-time at the first, second, or third quartile of the OP, or none. 
Given the lack of prior knowledge, we choose priors based on visualizing the corresponding hazard. Otherwise, we used the same model set-up as in the simulation study.

If the ESS is lower than $10\,000$ or the scale reduction factor\cite{stanmanual} $\hat{R}$ far from one, a larger posterior sample is generated. The BPgWSP test is applied with the recommended tuning of the simulation study.

The R code and results are stored on \url{https://osf.io/p8hvr/files/osfstorage}. See the supplementary material for a more detailed description of the analysis.

\subsection{Headache}
\begin{figure}[t]
    \centering
    \includegraphics[width = \textwidth]{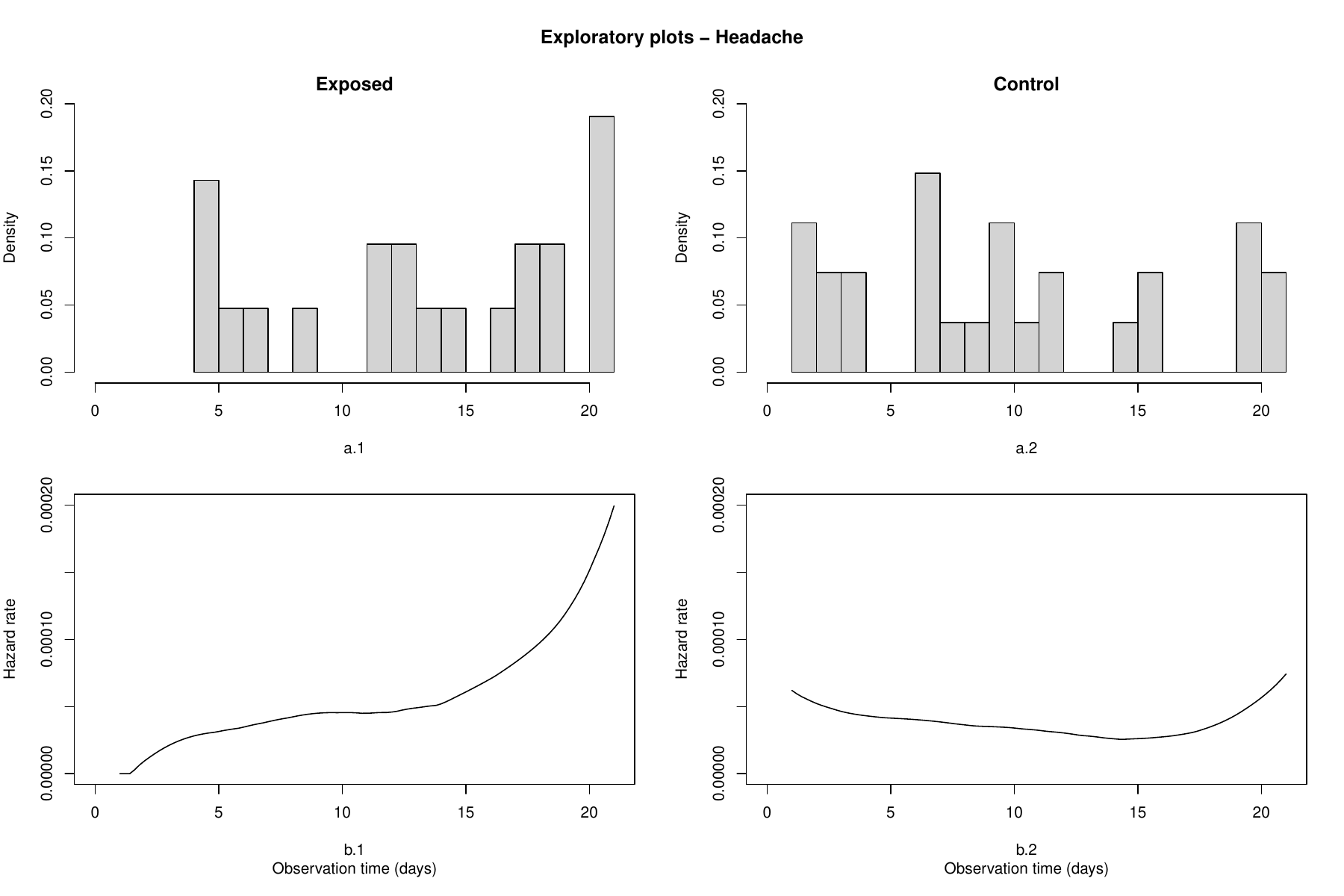}
    \caption{Graphics for exploratory analysis of the occurrence of headache over 21 days. The graphic derived is based on Sauzet et al., Fig. 1. \cite{sauzet2013illustration}}\label{fig:headache_exploratory_plots}
\end{figure}

\paragraph{Exploratory analysis:} The data is censored after 21 days. In this time, 21 among 19791 patients $(0.11\%)$ exposed to bisphosponates experienced headaches. The day of occurrence is on average $14 \; (sd = 5.9)$. The histogram of event-times (Figure \ref{fig:headache_exploratory_plots} a.1) shows that the number of events is slightly higher in the second half of the OP.
The non-parametric smooth hazard estimate (Figure \ref{fig:headache_exploratory_plots} b.1) indicates a more pronounced increase of risk in the last seven days.

Within the control group, headache occurred in 27 among 39630 patients $(0.07\%)$. The events are distributed over the full observation window (see Figure \ref{fig:headache_exploratory_plots} a.2). The hazard estimate is almost horizontal 
(see Figure \ref{fig:headache_exploratory_plots} b.2). 

\begin{table}[h]
    \centering
    \begin{tabular}{|l|r|c|c|c|}
        \hline
        prior belief about & a priori $E[t]$ & prior means  & test outcome & test outcome \\
        expected event-time & & $(\bar{\theta}_0, \bar{\nu}_0, \bar{\gamma}_0)$ & (Exposed) & (Control) \\
        \hline
        none & - & $(1, 1, 1)$ & - & -  \\
        1st quarter of OP & 5.25 days & $(2.5, 0.5, 1)$ & - & - \\
        2nd quarter of OP & 10.5 days & $(3, 3, 5)$ & signal &  -\\
        3rd quarter of OP & 15.75 days & $(18, 5, 1)$ & signal & - \\
        \hline
    \end{tabular}
    \caption{Test outcomes under various prior beliefs for headache based on exposed and control group. Prior standard deviation is set to 10 for all parameters in all prior belief cases.}
    \label{tab:headache_testoutcome}
\end{table}

\paragraph{Signal detection test:} A signal is raised for the exposed group under the prior belief that ADR occurs towards the middle or end of the OP. No signal is raised under the prior of no ADR or ADR at the beginning of the OP (see Table \ref{tab:headache_testoutcome}).
For the control group, no signal is raised independent of the prior specification.

Headache is a known ADR of bisphosphonates. \cite{bisphosphonatesAEreview}
The BPgWSP test detects a signal under the prior belief of ADR occurrence towards the middle or end of the OP. These beliefs presumably amplify the signal in the data already indicating a higher hazard rate towards the end of the OP (see Figure \ref{fig:headache_exploratory_plots} b.1) while the other prior settings might fail to detect a signal due to the contradictory prior and data information.

\subsection{Musculoskeletal pain}

\begin{figure}[t]
    \centering
    \includegraphics[width = \textwidth]{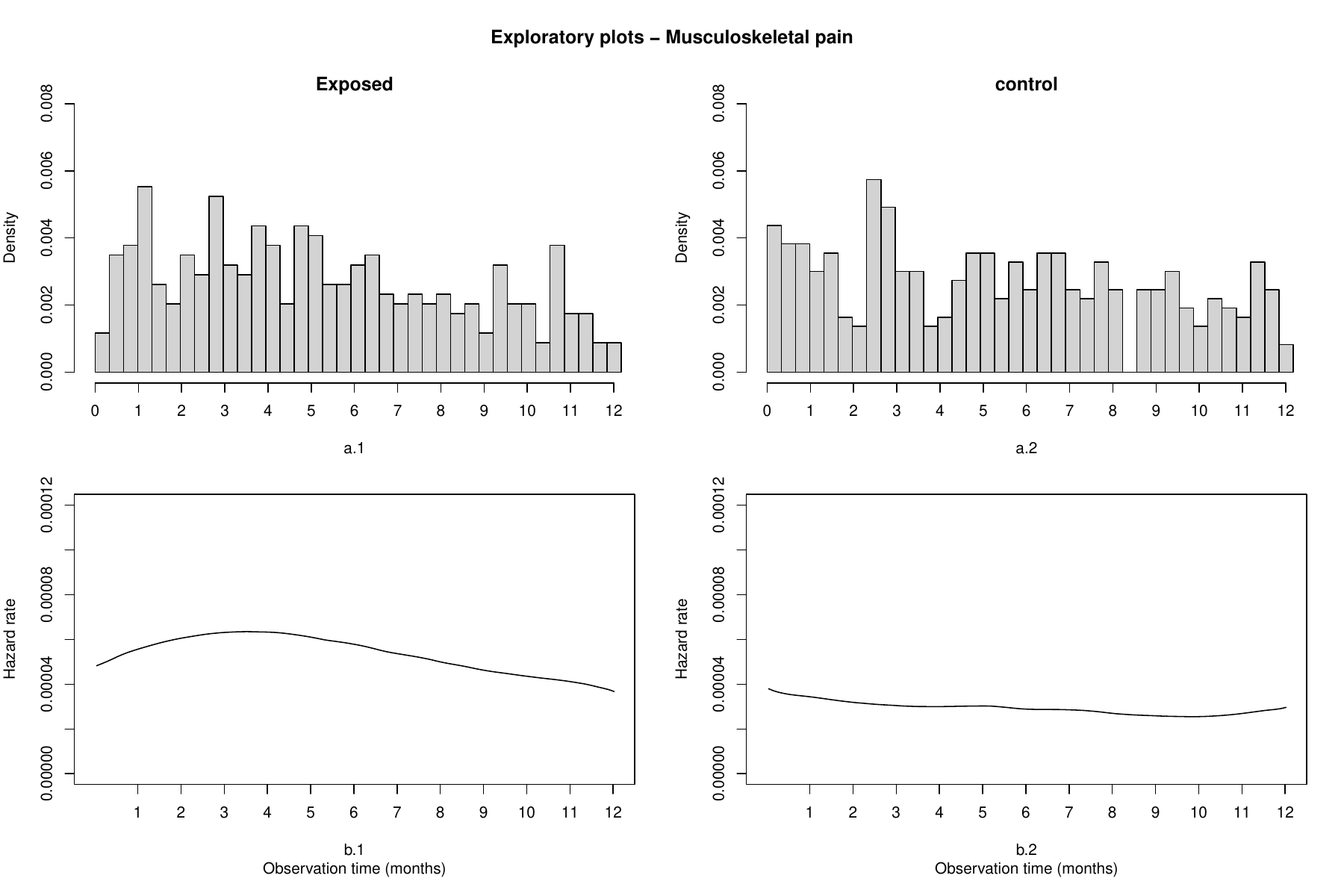}
    \caption{Graphics for exploratory analysis of the occurrence of musculoskeletal pain over one year. The graphic derived is based on Sauzet et al., Fig. 2. \cite{sauzet2013illustration}}\label{fig:musco_exploratory_plots}
\end{figure}

\paragraph{Exploratory analysis:}
The data is censored after one year in which 344 of 19777 ($1.7\%$) exposed patients reported musculoskeletal pain. The average event-time is 160 days $(sd = 99)$.
The histogram of event-times (Figure \ref{fig:musco_exploratory_plots} a.1) shows a slightly higher density in the first half of the OP. The smoothed hazard estimate (Figure \ref{fig:musco_exploratory_plots} b.1) indicates a slight increase in risk during the first four months followed by a decrease.

In the control group musculoskeletal pain was reported 366 times from 39634 patients ($0.9\%$). According to the histogram (Figure \ref{fig:musco_exploratory_plots} a.2) slightly more cases are reported in the first and third month and evenly distributed for the rest of the time. The hazard estimate (Figure \ref{fig:musco_exploratory_plots} b.2) looks almost constant.

\begin{table}[h]
    \centering
    \begin{tabular}{|l|r|c|c|c|}
        \hline
        prior belief about & a priori $E[t]$ & prior means  & test outcome & test outcome \\
        expected event-time & & $(\bar{\theta}_0, \bar{\nu}_0, \bar{\gamma}_0)$ & (Exposed) & (Control) \\
        \hline
        none & - & $(1, 1, 1)$ & signal & -  \\
        1st quarter of OP & 91 days & $(1, 0.207, 1)$ & signal & -  \\
        2nd quarter of OP & 183 days & $(20, 5.5, 14)$ & signal &  -\\
        3rd quarter of OP & 274 days & $(300, 4, 1)$ & - & - \\
        \hline
    \end{tabular}
    \caption{Test outcomes under various prior beliefs for musculoskeletal pain based on exposed and control group. Prior standard deviation is set to 10 for all parameters in all prior belief cases.}
    \label{tab:musco_testoutcome}
\end{table}

\paragraph{Signal detection test:} A signal is raised for the exposed group under all priors except a prior belief around the third quarter (day 274).
For the control group, no signal is raised regardless of the prior choice (see Table \ref{tab:musco_testoutcome}).

Musculoskeletal pain is a recognized ADR that can occur anytime after the beginning of bisphosphonate intake.\cite{adr_bis_muscu}
Here, the test result depends on the prior specification. A long OP and a small number of cases potentially decrease the performance of the BPgWSP test when the prior contrasts sharply with the data, as is the case under the end-prior here. Moreover, the result of no signal for the exposed group that was based on the model that was not rerun might indicate the need for more iterations, even when $\hat{R}$ and ESS suggest convergence.

\subsection{Alopecia}

\begin{figure}[t]
    \centering
    \includegraphics[width = \textwidth]{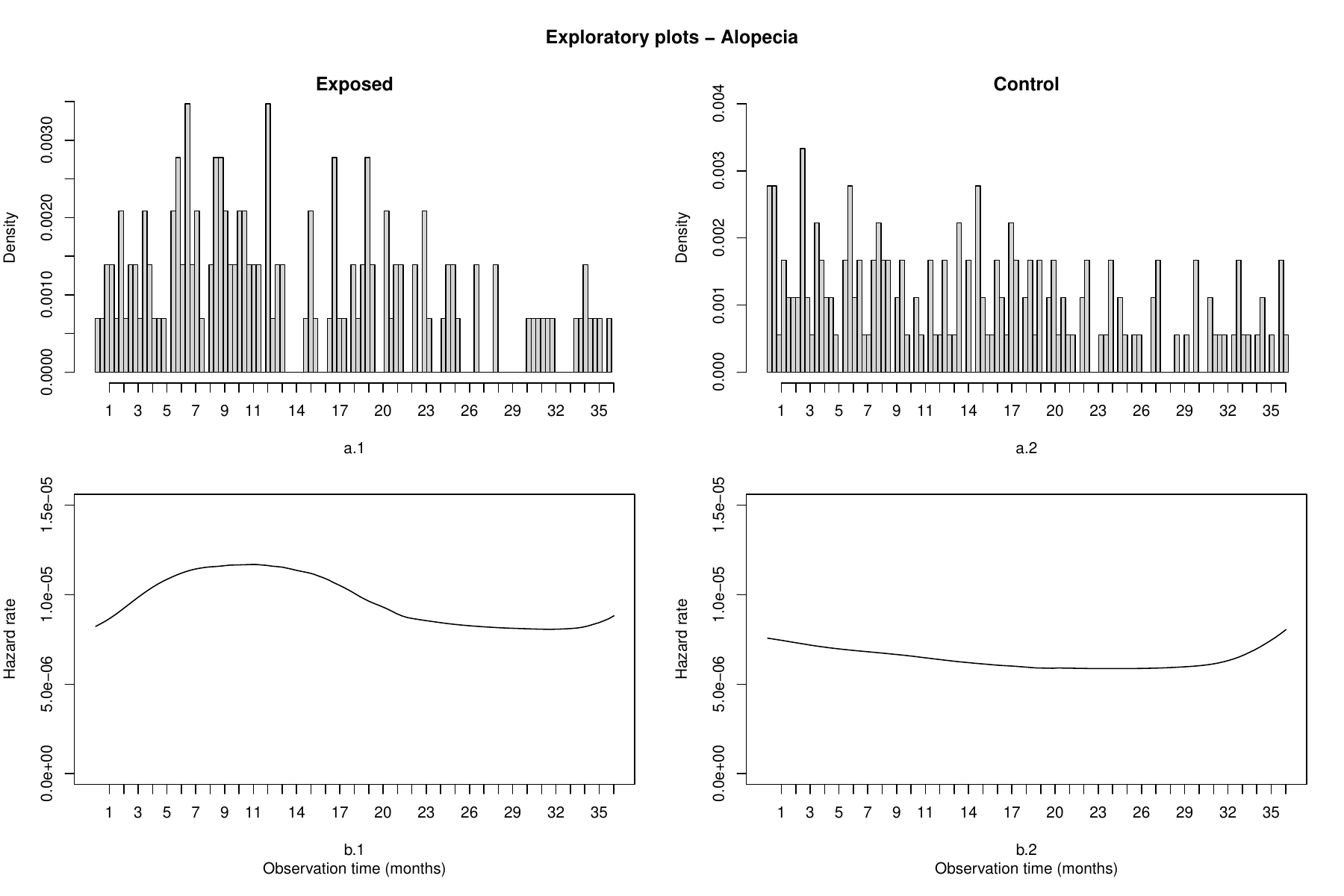}
    \caption{Graphics for exploratory analysis of the occurrence of alopecia over three years. The graphic derived is based on Sauzet et al., Fig. 3. \cite{sauzet2013illustration}}\label{fig:alopecia_exploratory_plots}
\end{figure}

\paragraph{Exploratory analysis:}
Over three years, there have been 144 reports of alopecia in the group of 19812 patients (0.73\%) exposed to bisphosphonates. The mean event-time is day 424 $(sd = 285)$. According to the histogram (Figure \ref{fig:alopecia_exploratory_plots} 1.a), events take place slightly more often between the fifth and 13th month which roughly corresponds to the period around the first or second quarter of the OP. Regarding the estimated hazard function, the risk rises and falls within the first two years and stabilizes after that.

There are 180 alopecia cases among 39634 control patients ($0.45\%$). The event-times are slightly denser in the first half of the observation window. The estimated hazard (Figure \ref{fig:alopecia_exploratory_plots} 2.b) has an almost flat trend.

\begin{table}[h]
    \centering
    \begin{tabular}{|l|r|c|c|c|}
        \hline
        prior belief about & a priori $E[t]$ & prior means  & test outcome & test outcome \\
        expected event-time & & $(\bar{\theta}_0, \bar{\nu}_0, \bar{\gamma}_0)$ & (Exposed) & (Control) \\
        \hline
        none & - & $(1, 1, 1)$ & signal & -  \\
        1st quarter of OP & 274 days & $(200, 0.63, 1)$ & signal & -  \\
        2nd quarter of OP & 548 days & $(30, 3.2, 10)$ & signal &  -\\
        3rd quarter of OP & 821 days & $(700, 14, 4)$ & - & - \\
        \hline
    \end{tabular}
    \caption{Test results under various prior beliefs for alopecia based on exposed and control group. Prior standard deviation is set to 10 for all parameters in all prior belief cases.}
    \label{tab:alopecia_testoutcome}
\end{table}

\paragraph{Signal detection test:}

Under the prior assumptions of no association, ADR occurrence in the beginning or in the middle of the OP the test identifies a temporal pattern for the exposed group and none for the control group. For both groups, no signal is raised under the prior belief that alopecia risk increases at the end of the OP (see Table \ref{tab:alopecia_testoutcome}).

Alopecia is a rare event not certainly identified as an ADR of bisphosphonates at the current state of research to our knowledge. The same uncertainty is reflected in the BPgWSP test results. Under an ADR-suspecting prior, no signal is raised for the control group and a signal flagged for the exposed group indicating that the signal might not be raised due to noise. Even with larger posterior samples under the none-prior, no signal was identified. This indicates that more iterations do not necessarily lead to a higher signal rate.

\subsection{Carpal tunnel syndrome}

\begin{figure}[t]
    \centering
    \includegraphics[width = \textwidth]{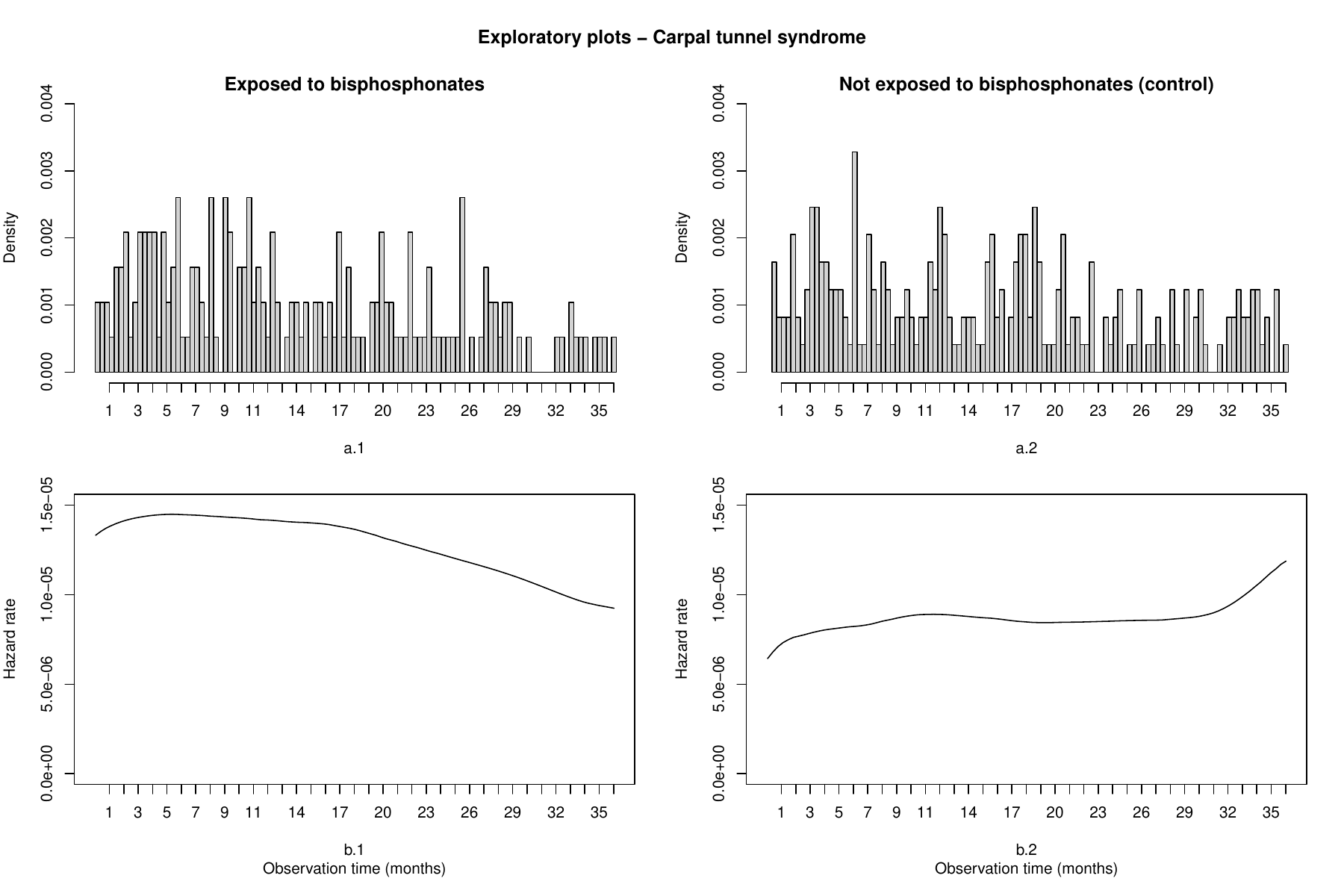}
    \caption{Graphics for exploratory analysis of the occurrence of carpal tunnel syndrome over three years. The graphic derived is based on Sauzet et al., Fig. 4. \cite{sauzet2013illustration}}\label{fig:tunnel_exploratory_plots}
\end{figure}

\paragraph{Exploratory Analysis:}
Within the OP of three years, 192 among 19817 exposed patients ($0.97\%$) experienced carpal tunnel syndrome. The day of occurrence is on average $429 \; (sd = 290)$ which is approximately one year after drug prescription. The histogram of event-times (Figure \ref{fig:tunnel_exploratory_plots} a.1) reveals that events very slightly decrease over the scope of the OP.
The hazard estimate (Figure \ref{fig:tunnel_exploratory_plots} b.1) indicates a slightly higher constant risk in the first 18 months and a decrease after that. 

Within the control group, carpal tunnel syndrome occurred for 244 among 39389 patients ($0.62\%$). The distribution of events looks similar to the exposed group's plot (see Figure \ref{fig:headache_exploratory_plots} a.2). The hazard estimate shows an incline of risk in the first year, a rather constant trend in the middle and again an increase within the last 5 months. (see Figure \ref{fig:tunnel_exploratory_plots} b.2).

\paragraph{Signal detection test:}

In two out of four prior specifications, namely when the prior belief is that the event is to be expected around the first quarter or middle of the OP, a signal is raised for both the control and exposed group. In the two other cases, a signal is raised for neither group (see Table \ref{tab:tunnel_testoutcome}).

Carpal tunnel syndrome is not a recognized ADR to our knowledge.
The ambiguous test results assigning signals to exposed and control group under a subset of prior specifications are probably returned due to the confounding effect of the age of the participants for which the carpal tunnel syndrome is a common condition.
This suggests that sample stratification might be an important preparation before BPgWSP testing in some applications.
\begin{table}[h!]
    \centering
    \begin{tabular}{|l|r|c|c|c|}
        \hline
        prior belief about & a priori $E[t]$ & prior means  & test outcome & test outcome \\
        expected event-time & & $(\bar{\theta}_0, \bar{\nu}_0, \bar{\gamma}_0)$ & (Exposed) & (Control) \\
        \hline
        none & - & $(1, 1, 1)$ & - & -  \\
        1st quarter of OP & 274 days & $(200, 0.63, 1)$ & signal &  signal\\
        2nd quarter of OP & 548 days & $(30, 3.2, 10)$ & signal &   signal\\
        3rd quarter of OP & 821 days & $(700, 14, 4)$ & - & - \\
        \hline
    \end{tabular}
    \caption{Test outcomes under various prior beliefs for carpal tunnel syndrome based on exposed and control group. Prior standard deviation is set to 10 for all parameters in all prior belief cases.}
    \label{tab:tunnel_testoutcome}
\end{table}

\section{Discussion}

In this work we developed the BPgWSP test that raises a signal if there is an indication of non-constant hazard over time for an AE after drug intake incorporating prior knowledge about the time of occurrence. The defined test was tuned based on a simulation study which led to a one-fits-all approach. 
\\

An important aspect of our proposed BPgWSP test is the role of prior information in shaping its performance. The BPgWSP test performs very well under correct prior assumptions and outperforms the frequentist approach. If the prior specification is wrong, the performance of the test decreases considerably which is also reflected in the case study. 

We recommend using the BPgWSP test only in case of certainty about prior information, and otherwise, use the frequentist approach.
In light of these results, correct prior elicitation is of importance. It is advised to follow procedures recommended by others. See O{'}Hagan for an introduction.\cite{prior_el} 
\\

The simulation study enabled an initial assessment of the test performance and the conclusion of a recommended BPgWSP test tuning for application in an automated one-fits-all manner to a multitude of drug-AE combinations.  
In general, however, we recommend tuning the BPgWSP test with an additional simulation study. 
For this purpose, we developed the R package BWSPsignal.\cite{BWSPsignal}

\section*{Data availability statement}
Simulation study output as well as the R package BWSPsignal and R scripts for reproducibility are available on \url{https://github.com/julia-dyck/BWSPsignal}.

The data used in the case study section was originally provided from the THIN \cite{THINwebsite} database. Please contact the corresponding author upon reasonable request.

\section*{Funding}
No funding was received to perform this study.

\section*{Conflicts of interest}
The authors declare no conflict of interest.

\newpage
\bibliographystyle{ama} 

\bibliography{badrrefs}

\end{document}